\documentclass[letterpaper, reqno]{amsart}
\usepackage{epsfig}
\usepackage{graphicx}
\usepackage{oldlfont}
\usepackage{amssymb}

\newtheorem*{ack}{Acknowledgment}
\theoremstyle{plain}

\newcommand{\ket}[1]{| #1 \rangle}

\newcommand{\eqngraph}[2][1]{\mbox{
\raisebox{-0.5\totalheight}{\includegraphics[scale=.65]{#2.eps}}}}

\begin{document}

\title{Modelling Space with an Atom of Quantum Geometry}

\author{Seth A. Major and Michael D. Seifert}

\date{August 2001}
\address{Major: Department of Physics\\
Hamilton College\\
Clinton NY 13323 USA\\Seifert: Department of Physics and Astronomy\\
Swarthmore College\\
Swarthmore PA 19081 USA}
\email{Major: smajor@hamilton.edu \\ Seifert: seifert@sccs.swarthmore.edu}

\begin{abstract}
Within the context of loop quantum gravity there are several operators
which measure geometry quantities.  This work examines two of these
operators, volume and angle, to study quantum geometry at a single spin
network vertex - ``an atom of geometry.''  Several aspects of the angle
operator are examined in detail including minimum angles, level spacing,
and the distribution of angles.  The high spin limit of the volume
operator is also studied for monochromatic vertices.  The results show
that demands of the correct scaling relations between area and volume and
requirements of the expected behavior of angles in three dimensional flat
space require high-valence vertices with total spins of approximately
$10^{20}$.

\end{abstract}

\maketitle

\section{Introduction}
\label{intro}

Several promising theories of quantum gravity predict that space is
discrete.  Within the context of loop quantum gravity, operators for
area \cite{RSareavol,FLR,QGI}, volume \cite{RSareavol,QGII,RC}, and
possibly even angle \cite{angle} have discrete spectra.  Given this
radical departure from the paradigm of continuous space that has
served us so well, it is important to study consequences of the suggested
underlying discrete nature of space.  This paper reports on a study on the
angle and volume operators.  More extensive discussion may be found in Ref.
\cite{mikethesis}.  Some earlier results were reported in Ref. \cite{mg9}.

The eigenbasis for all geometric operators is the spin network basis. 
In this context spin networks are not merely a graphical notation for
representation theory but also contain information that determines the
properties of space.  All geometric and matter degrees of freedom are
contained in the network.  Indeed, as traditional spin networks are
embedded in a three dimensional space knots in the network contain
physical information about the state of quantum geometry.

Geometric quantities are concentrated in the edges and vertices of the
spin network.  If an edge intersects a surface, the edge endows that
surface with area.  If a vertex (with valence larger than 3) is
contained within a region, then the spin network vertex contributes
volume to that region.  The angle operator also is defined at a single
spin network vertex.  Since volume and angle are located at vertices,
the spin network vertex may be called an ``atom of geometry.''  In
addition if we look beyond the kinematic level, most of the current
proposals to implement the Hamiltonian constraint contain operator
expressions which act at the spin network vertex.

This work is a first step in what we call ``single vertex studies.''
Motivated by the fact that volume and angle operators act at a single
vertex and the central importance of the spin network vertex in dynamics,
we studied the relation on the geometric quantities defined on this space.
 In such a simple context we can answer such questions as: ``What are the
properties of spin network states which approximate continuous space?'' 
and, ``What are some of the surprises that arise out of this new quantum
geometry?''

In the next section we give the necessary notation, a very brief
review of the angle operator, and the main analysis.  In Section
\ref{volres} we do the same for the volume operator.  Section
\ref{conc} contains a summary of the key points of the study and some
additional consequences.

\section{Angle Results}

\begin{figure}
      \begin{center}
  \begin{tabular*}{\textwidth}{c@{\extracolsep{\fill}}
  c@{\extracolsep{\fill}}}
  \includegraphics[scale=.75]{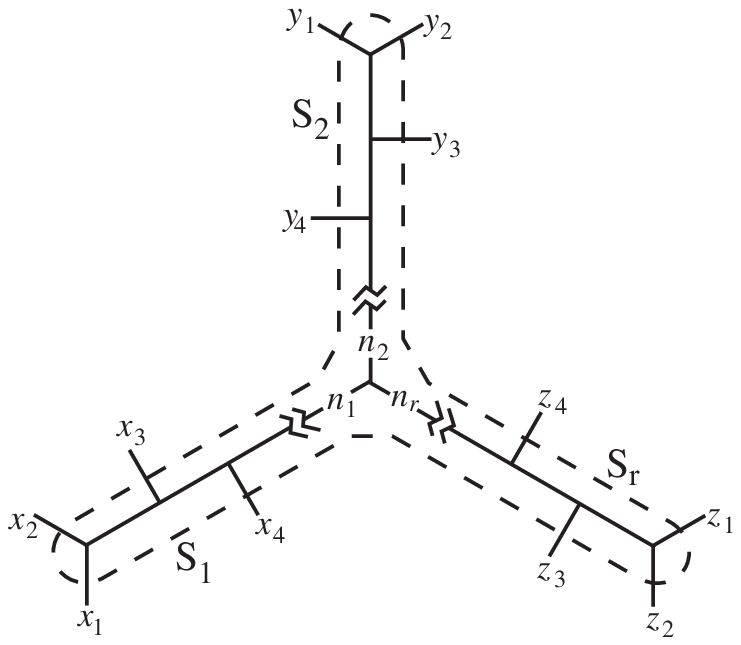}
    &
  \includegraphics[scale=.85]{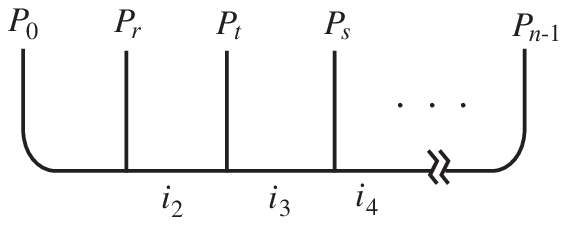}\\
  (a.) &  (b.)
  \end{tabular*}
  \end{center}
      \caption{\label{ints} Two useful intertwiners.  (a) Internal
      decomposition of a vertex operated on by the angle operator; the
      dotted line denotes the boundary between external and internal
      edges.  All of the edges $x_1, x_2, \ldots$ which intersect the
      surface $S_1$ are ``collected'' into the internal edge labelled
      with
      $n_1$; similarly, all the edges which intersect $S_2$ are collected
      into the edge labelled $n_2$, and the edges which intersect $S_r$  are
      collected into the edge labelled $n_r$.  In the text, we denote $s_1
      = \sum x_i$, $s_2 = \sum y_i$, and $s_r = \sum z_i$.  This is called
      the snowflake basis.  (b) The external edges $P_0, P_1, \ldots,
      P_{n-1}$ and the internal edges $i_2, i_3, \ldots, i_{n-2}$ of the
      comb basis.}
\end{figure}

Before reporting on the single vertex study of the angle and volume
operators we need to introduce some notation on the spin network
vertex.  The number of incident edges is called the valence.  While the
internal structure of a trivalent vertex is unique, higher valence
vertices can be constructed in more than one way.  These different
internal structures are described by the intertwiner, which labels the
ways in which the incident representations are connected. Intertwiners
therefore label a basis for the vertex.

The intertwiner may be described by a set of trivalent vertices
connected by ``internal edges.''  When the edges are to be partitioned
into three categories, as in the case of the angle operator, we can define
a convenient basis in which the external edges of one category are
collected in a branch which ends in one principal, internal edge. The
three principal edges are then connected in a vertex we call the
``intertwiner core.''  This (class of) intertwiners is called the
``snowflake'' basis shown in Figure \ref{ints}a.  The core is the only
part of the intertwiner which must be specified before completing the
diagrammatic calculation of the angle spectrum.  When all incident edges
have the same label the vertex is called ``monochromatic.''

\subsection{Angle operator: Definition}
\label{anglerev}

The idea that one could measure angle from the discrete structure of
spin networks was first suggested by Penrose \cite{penrose}.  This
work culminated in the Spin Geometry Theorem, which stated that angles in
3-dimensional space could be approximated to arbitrary accuracy if the
spin network was sufficiently correlated and if the spins were
sufficiently large \cite{JM}.  This result may be seen as the ``high spin
limit'' of the observation that, for an EPR pair, we know the relative
orientation of the particles' spins but nothing about the absolute
orientation of the two spins in a background space.

\begin{figure}
       \begin{center}
      \includegraphics[scale=.85]{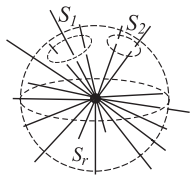}
       \end{center}
    \caption{ \label{cones} A vertex with three 
partitions of edges
    identified by the surfaces $S_{1}$, $S_{2}$, and $S_{r}$.
    The core of the intertwiner for the vertex is chosen so that
    all edges passing through $S_{i}$ connect to a single
    internal edge $n_{i}$ (with $i=1,2,r$) as shown in Figure
    \ref{ints}a.}
\end{figure}

The spin geometry construction carries over into the context of
quantum geometry: one can measure the angle between two internal edges
of a spin network state.  At the outset we note that there is more
than one definition of the angle operator \cite{angle}.  In this paper
we use the quantization that most closely matches the operator in the
spin geometry theorem.  This angle operator is a quantization of the
classical expression for an angle $\theta_{v}$ measured at a point
$v$.  The operator assigns an angle to two bundles of edges incident
to the vertex $v$.  These edges are identified by the two surfaces
$S_{1}$ and $S_{2}$ shown in Figure  \ref{cones}.  Likewise, there are
two angular momenta operators $J_{1}$ and $J_{2}$ associated to these
surfaces.  Acting on the spin network state $\ket{s}$ the angle
is \cite{angle}
\begin{equation}
      \label{aop}
  \hat{\theta}^{(12)}_{v} \ket{s} := \arccos \frac{\hat{J}_{1} \cdot
  \hat{J}_{2}} {|\hat{J}_{1}| \, | \hat{J}_{2} |} \ket{s}.
\end{equation}

All edges incident to $v$ are partitioned into three categories
corresponding to which surface they pass through.  In each category,
the edges connect to a single internal edge in the intertwiner core. 
These edges are denoted $n_{1}$, $n_{2}$, and $n_{r}$.  For reasons
that will be clear shortly, the spin $n_{r}$ is known as the ``geometric
support.''  Using the usual relations for angular momentum operators
and an intertwiner core labelled by $n_{1}$, $n_{2}$ and $n_{r}$, one
immediately finds the spectrum
  \begin{equation}
  \label{anglespec}
  \hat{\theta}^{(12)}_{v} \ket{s}
  = {\rm arccos} \left[ \frac{n_{r}(n_{r}+2) -
  n_{1}(n_{1} +2) - n_{2}(n_{2}+2)} {2 \sqrt{ n_{1}(n_{1} +2 ) \,
  n_{2}(n_{2} +2 ) } } \right] \ket{s}.
  \end{equation}
We use the integer labels for which $j_{i}= n_{i}/2$.  The key idea of
the angle operator is to measure the relative spins of internal edges
identified with the surfaces $S_{1}, S_{2}$, and $S_{r}$.

Note that the quantities $n_1$, $n_2$, and $n_r$ are the internal
edges that ``collect'' the spins from each of the three surface
patches. They are related to the total edge flux intersecting each
patch by the relations
  \begin{equation}
  n_1 \leq s_1\mbox{, }n_2 \leq s_2\mbox{, and }n_r \leq s_r
  \end{equation}
where $s_1$, $s_2$, and $s_r$ are the respective surface fluxes.

Normally the discreteness in geometric observables such as area is on the
Planck scale $\ell_P \sim 10^{-35}$ m.  However, angle operators have no
dependence on the scale of the theory $\ell_P$.  The result is {\it
purely} combinatorial!  This result provides a new test bed to investigate
quantum geometry.\footnote{There is an important caveat: Retaining the
diffeomorphism invariance of the classical theory in the construction of
the state space, there is, generically, a set of continuous parameters or
moduli space associated with higher valence vertices \cite{moduli}.  These
parameters contain information on the embedding of the spin network graph
in space.  On such a space the angle spectrum is highly degenerate.  It is
by no means clear whether this embedding information is physically
relevant \cite{grottthesis}.  Indeed, the state space of quantum gravity
may be described by abstract or non-embedded graphs.}

\begin{figure}
       \includegraphics[width=3.75in]{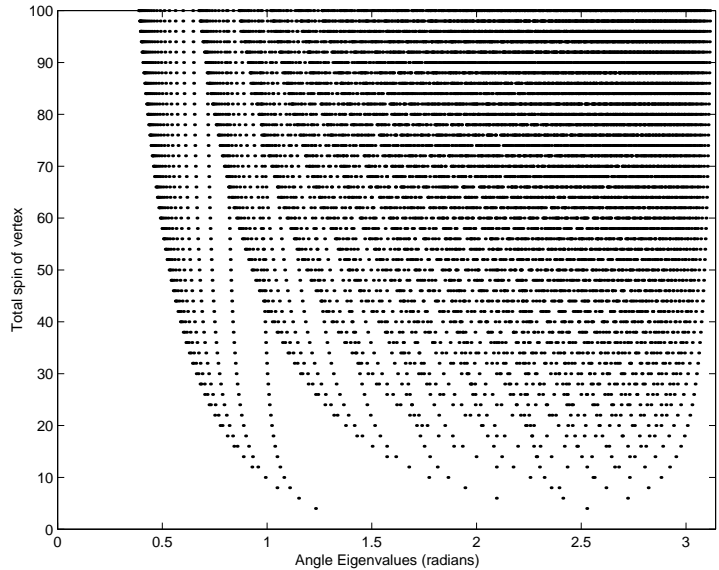}
   \caption{\label{AngSpect} Angle operator spectrum for increasing
   total vertex spin $n$.  The complete spectrum is plotted for
   total spins from 3 to 100.}
\end{figure}

Two aspects of the spectrum are immediately obvious from the spectra
in Figure \ref{AngSpect}.  First, the smallest angle can be quite
large; it is ``hard'' to model small angles.  Second, the gap between
neighboring angles is non-uniform.  It is clear from a glance that in
order to model continuous space both of these effects must be minimized. 
In addition to these more obvious aspects, there is a requirement of these
angles as embedded in three dimensional space on the distribution of
angles.  Each of these aspects are discussed in the following subsections.

\subsection{Small Angles}

To obtain a small angle, we wish for the argument of the inverse
cosine in (\ref{anglespec}),
\begin{equation}
  Q \equiv \frac{n_r (n_r + 2) - n_1 (n_1 + 2) - n_2 (n_2 + 2)}
  {2 \sqrt{n_1(n_1 + 2)n_2(n_2+2)}},
\end{equation}
to be as close to 1 as possible.  However, the internal edge spins are
subject to the triangle inequalities $$
  2n_1, 2n_2, 2n_r \leq n_1 + n_2 + n_r.
$$
To maximize $Q$, then, we wish to maximize its numerator and minimize its
denominator while simultaneously respecting the triangle inequalities. 
This occurs when \begin{equation}
  n_r = \frac{n}{2} \mbox{ and } n_1 = n_2 \approx \frac{n}{4}
\end{equation}
where the ``spin sum'' $n$ is defined as $n =
n_{1}+n_{2}+n_{r}$.\footnote{Note that while the value for $n_r$ is
exact, the values of $n_1$ and $n_2$ are only approximate; this is
because $n$ must be even, but may not necessarily be divisible by 4,
and because $n_1$, $n_2$, and $n_r$ must be integers.  However, for
large values of $n$, the values of $n_1$ and $n_2$ that maximize $Q$
will approach $n/4$.} At these values of $n_1$, $n_2$, and $n_r$, the
argument of the inverse cosine then reduces to \begin{equation}
  Q = \frac{n}{n + 8}.
\end{equation}
For any angle $\epsilon$, then, we have the relation
$$
  \epsilon \geq \arccos \left( \frac{n}{n + 8} \right).
$$
For small $\epsilon$, $\cos x$ is a uniformly decreasing function;
hence, for $\epsilon \ll 1$, we have
\begin{equation}
      \label{SmallAng}
  \epsilon \geq \frac{4}{\sqrt{n + 8}}.
\end{equation}

From this relation we see that the minimum possible angle between two
patches is proportional to $n^{-1/2}$ for large $n$.  Moreover, we
note that if $s_r > s_1 + s_2$, then $n \leq 2(s_1 + s_2)$ which
implies that for a ``macroscopic'' vertex, where $s_r \gg s_1, s_2$,
the minimum observable angle depends only on the edge flux through the
patches whose angular separation we are measuring.  Since the total
spin is bounded from above by twice the spin fluxes through the cones,
the minimum observable angle depends on the area flux through the
regions we are measuring.  For instance, to achieve an angle as small
as $10^{-10}$ radians, the required flux is roughly $10^{20}$.

\subsection{Mean Angular Resolution}
\label{angres}

The second property of the spectrum is the non-uniform level spacing. 
It is possible to quantify the angular resolution of a given vertex by
counting the number of possible results of an angle measurement.  We
define the mean angular resolution $\delta$ of a vertex to be
the average separation between each possible angle and the next
greatest possible angle.  In other words, $\delta$ is the average
width of a ``gap'' between all possible angles in the interval $[0,
\pi]$.  This quantity is not the angular resolution for a given angle
measurement; we do not compute the angle gap given an angle
eigenvalue.

The simplest way of finding the mean angular resolution for a given
vertex is to note that an $n$-valent vertex has $n-2$ trivalent
vertices in its intertwiner.  Since each one of these internal
vertices can be looked at as a core vertex, we can conclude that such
a vertex has at most $3(n-2)$ possible eigenvalues for the angle
operator.  Since the $3n-6$ eigenvalues associated with the vertex
divide the interval into $3n-5$ gaps, we conclude that the mean angular
separation between the eigenvalues of the angle operator for an
$n$-valent vertex is
\[
\delta \geq  \frac{\pi}{3n-5}.
\]
The inequality accounts for possible degeneracies due to the fact that
internal vertices may yield the same angles.
 
This method is limited.  It only examines the separation between the
eigenvalues of the angle operators that have a given vertex as their
eigenstate.  However, other angle measurements are possible for a
given vertex.  For example, if we measure the angle between two edges
that do not immediately intersect in the vertex's intertwiner, this
measurement will not always yield a definite eigenvalue.  Instead,
there will be an expectation value associated with this measurement. 
The question then becomes: for a given vertex, can we find the total
number of possible expectation values?
 
To measure an angle we must partition the edges into three groups (one
each for $S_{1}, S_{2}$, and $S_{r}$).  In the snowflake basis, we
will generally obtain a superposition of angle eigenstates. 
Nonetheless, given a partition we will measure a single, definite
expectation value, despite the superposition of eigenvalues.  We
conclude that the number of possible expectation values is equal to
the number of distinct partitions of the edges.
 
Since the possible range of colours for the collecting edge is
not dependent on the order of the edges along the branch, the
number of distinct partitions is given by
\begin{equation}
 {\mathcal N} = \prod_{i=1}^{\infty} B(q_i) \mbox{,}
\end{equation}
where $q_i$ is the number of edges with colour $i$, and $B(n)$ is the
number of distinct ways of partitioning $n$ into three distinct bins. 
Elementary combinatorics tells us that the number of ways to do this
is
\begin{equation}
 B(n) = {(n + 3) - 1 \choose n} = \frac{(n+1)(n+2)}{2}.
\end{equation}
We must divide the overall product by two, since exchange of the sets
corresponding to the $n_1$ branch and the $n_2$ branch yields the same
angle.  Thus, the maximum number of possible expectation values for a
given vertex is
\begin{equation}
 {\mathcal N} = \frac{1}{2} \prod_{i=1}^{\infty} 
 \frac{(q_i+1)(q_i+2)}{2} \mbox{.}
\end{equation}
In the monochromatic case, where all the edges have the same colour,
$q_i$ is equal to the vertex valence $n$ for one value of $i$ (the
edge colour) and zero otherwise.  The formula reduces to
\begin{equation}
 {\mathcal N} = \frac{(n+1)(n+2)}{4}
\end{equation}
and the mean angular resolution (between expectation values) is given
by
\begin{equation}
 \delta_{v} \geq \frac{4 \pi}{n^2 + 3n + 3}
\end{equation}
where we have again used an inequality because of possible angle
degeneracies.
 
It is important not to confuse the ``mean angular resolution'' as we
have defined it with the average distance from a random point in the
interval $[0, \pi]$ to the nearest angle.  What we have found allows
us to get an idea of how tightly, on average, the eigenvalues of the
angle operator for a given vertex are spaced.  To find this other
value (which, if found, would help us to estimate the level to which
an arbitrary angle can be approximated) we would need
\begin{equation}
\langle \Delta \rangle = \sum_{\mathrm{all \: gaps}} {\mbox{prob.  of 
being} \choose \mbox{located in the gap}} {\mbox{mean separation 
from} \choose \mbox{either end of gap}}.
\end{equation}
This is equivalent to the expression
\begin{equation}
 \langle \Delta \rangle = \sum_{a} \frac{w_a^2}{4\pi}
\end{equation}
where $w_a$ is the width of the gap $a$.  This expression depends
heavily on the precise location of the eigenvalues in the interval
$[0, \pi]$, and is therefore difficult to compute analytically.

\subsection{Angle Distribution}

In the classical continuum model of 3-dimensional space, the
distribution of solid angles is proportional to $\sin \theta$
\begin{equation}
  \mathcal{P}(\theta) \, \mathrm{d} \theta = \sin \theta \, \mathrm{d}
  \theta .
\end{equation}
If our quantized angle operator is to be viable, it must reproduce
this distribution in some ``classical limit.''  To find out whether
this is so, we must examine not only the location of possible
eigenvalues in the interval $[0, \pi]$, as we have in the past two
subsections, but also the likelihood with which these angles occur.
Due to the complexity of this problem, we will consider only the
simplest possible case which could still conceivably have a classical
limit: that of an $n$-valent monochromatic vertex with edge colour 1.

We note that every vertex can be transformed into the snowflake basis by
repeated application of the recoupling theorem.  For a random vertex,
however, we do not know the exact internal structure \emph{a priori};
there is no reason to assume that any given intertwiner is
preferred.  We will therefore assume that all intertwiners are equally
likely.  With this simplification in mind, we can conclude that the
probability of measuring the angle associated with an intertwiner core
$n_1, n_2, n_r$ is given by \begin{equation}
  {\mathcal P} (\theta(n_1, n_2, n_r)) \propto {\mbox{number of
  intertwiners} \choose \mbox{with core } n_1, n_2, n_r}.
\end{equation}

We wish to know, then, how many intertwiners exist with a given $n_1$,
$n_2$, and $n_r$.  It can be shown that if $i$ is the number of edges (of
colour 1) entering a branch, the number of distinct branches $\{a_2,
a_{3}, \ldots, a_{i}\}$ that end in a given $n_{1} = a_{i}$ is
\cite{mikethesis} \begin{equation} \label{collnum}
  Q(i, a_{i}) = \frac{i+1}{a_{i}+1} {a_{i}+1 \choose
  \frac{a_{i}+i}{2} + 1}
\end{equation}
To turn this into a normalized probability distribution, we
approximate the binomial coefficient in the above expression by a
Gaussian distribution multiplied by a normalization factor
\[
  {x \choose y} = A \exp \left[
  \frac{-(x-\frac{y}{2})^{2}}{\frac{y}{2}} \right],
\]
where $A$ is a normalization factor.  Applying this to
(\ref{collnum}), we have
\begin{equation}
  P(i, a_{i}) = \frac{i+1}{a_{i}+1} \exp \left[ \frac{-(a_i^{2} +
  2 a_i)}{2(i+1)} \right].
\end{equation}
$A$ has been subsumed into the exponential term.

\begin{figure}[p]
    \begin{center}
    \includegraphics[width=4in]{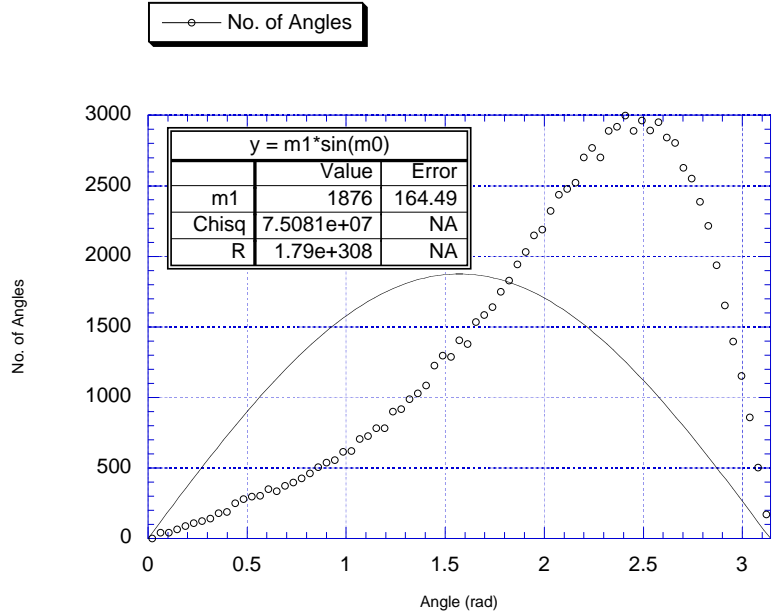}
 \caption{\label{dist1} Angle distribution for $s_{1} = s_{2} =
 s_{r}$.}
     \end{center}
\end{figure}

\begin{figure}[p]
     \begin{center}
     \includegraphics[width=4in]{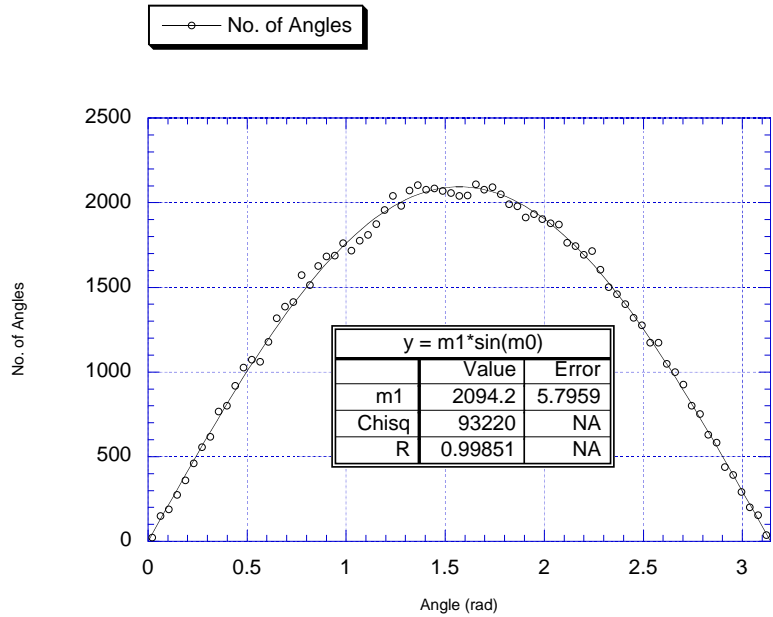}
 \caption{\label{dist3} Angle distribution for $s_{1} < s_{2} \ll
 s_{r}$.}
     \end{center}
\end{figure}

Using this probability, we can numerically compute the distribution of
the angles in the interval $[0, \pi]$.  However, the sheer number of
possible vertices for a macroscopic vertex prohibits calculating every
angle associated with such a vertex.  Instead, we sampled the possible
angles randomly, resulting in a certain amount of scatter in the data
points.  This allowed us to gain a qualitative understanding of the
distribution.  The results for various values of $s_{1}$, $s_{2}$, and
$s_{r}$ are shown in Figures (\ref{dist1}) and (\ref{dist3}). 

Figure (\ref{dist1}) shows the distribution for $s_{1} = s_{2} =
s_{r}$.  This distribution certainly does not match the expected
classical distribution; most notable is the fact that the distribution
peaks at approximately $2.5 \mbox{ rad} \approx 140^{\mathrm{o}}$. 
Figure (\ref{dist3}) shows the distribution for $s_{1} < s_{2} \ll
s_{r}$.  This distribution has surprisingly good correspondence with
the expected classical angle distribution.

These results, while not conclusive, are certainly indicative that the
angle operator can indeed reproduce the classical angle distribution in
the case where $s_{1}, s_{2} \gg 1$ and $s_{1}, s_{2} \ll s_{r}$. This
first condition corresponds to what normally thinks of as a ``classical
limit,'' well away from the regions where quantum effects dominate.  The
second condition can be thought of as a requirement for a sufficient
amount of ``background geometry'': not only do we need to make sure that
the angles we are measuring include a large amount of spin, but we also
need to ensure that the background spacetime upon which we measure the
angles is sufficiently classical.

\section{Volume Results}
  \label{volres}

\subsection{Volume Operator: Definition}
  \label{volrev}

After briefly providing the definition of the Rovelli-Smolin-DePietri
volume operator, we present the large spin results.  For more on the
operator itself, see Ref.  \cite{RC}.

The volume operator acts on a vertex.  In this case, however, it is
more convenient to use the basis pictured in Figure \ref{ints}b, where all
the edges are lined up sequentially.  One also defines the volume-squared
operator (known as $W$) first.  The volume eigenvalues are the square
roots of the absolute values of the eigenvalues of $W$.

For a vertex with $n$ edges labelled $\{P_{0}, P_{1}, \ldots,
P_{n-1}\}$, we define an operator $W_{[rst]}$ for each triple of edges
such that $0\leq r<s<t\leq n-1$.  $W_{[rst]}$ grasps edges $r$, $s$, and
$t$: \begin{equation}
  \hat{W}_{[rst]} \eqngraph{Combi}  = P_{r} P_{s} P_{t} \eqngraph{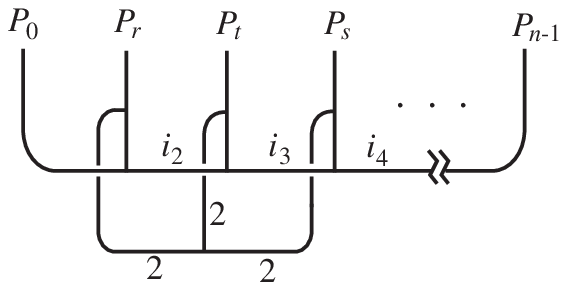}.
\end{equation}
We can write the left hand side of this equation as a superposition of
other vertices with the same internal decomposition as those in Figure
\ref{ints}b 
\begin{equation} \hat{W}_{[rst]} \eqngraph{Combi} =
\sum_{k_{2}, \ldots, k_{n-2}} W_{[rst]} {}_{i_{2}, \ldots,
i_{n-2}}^{k_{2}, \ldots, k_{n-2}} \eqngraph{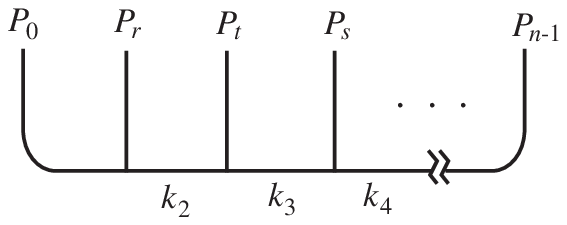}.
\end{equation}
This is essentially a matrix equation with the entries of the matrix
indexed by the internal intertwiner edges.  These entries turn out to be
\cite{RC} 
\begin{eqnarray}
      \label{Weq3}
  W^{(n)}_{[rst]} {}^{k_2 \cdots k_{n-2}}_{i_2 \cdots i_{n-2}} & = & - P_r
  P_s P_t \left\{ \begin{array}{ccc} k_2 & P_t & k_3 \\ i_2 & P_t & i_3 \\
  2 & 2 & 2 \end{array} \right\}  \lambda_{k_2}^{i_2 2} \delta_{i_4}^{k_4}
  \cdots \delta_{i_{n-2}}^{k_{n-2}} \nonumber \\ & & \, \times \frac{
  \left\{ \begin{array}{ccc} P_r & P_r & P_0 \\ k_2 & i_2 & 2 \end{array}
  \right\} \left\{ \begin{array}{ccc} P_s & P_s & k_4 \\ k_3 & i_3 & 2
  \end{array} \right\} }{\theta(k_2, k_3, P_t)}. \label{Weq2}
  \end{eqnarray} In this formula we have used the 9-$j$ symbol, which is
  equivalent to the spin network 
  \begin{equation} \left\{ 
  \begin{array}{ccc} k_2 & P_t & k_3 \\ i_2 & P_t & i_3 \\ 2 & 2 & 2
  \end{array} \right\} =
  \mbox{\raisebox{-1cm}{\includegraphics[totalheight=2cm]{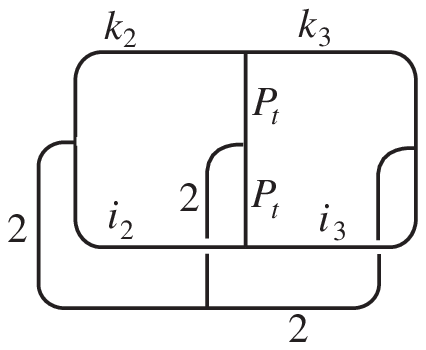}}}.
  \label{9jdef}
\end{equation}

The eigenvalues of the volume-squared operator are then proportional
to the sum of the absolute values of the eigenvalues of the
$W$-operators
\begin{equation}
      \label{voldef}
  V^2 = \sum_{0\leq r < s < t \leq n-1} \left| \frac{i}{16}
  W_{[rst]} \right|,
\end{equation}
and the volume eigenvalues are the square roots of the eigenvalues of
the $V^2$ operator.

\setlength{\arraycolsep}{1mm}
\subsection{The $W$-matrix and Eigenvalue Bounds}

One can see that the expression for the volume of Eq.~(\ref{voldef})
is horrendously difficult to handle in all its generality, for three
main reasons.  First, for an $n$-valent vertex with arbitrary edge
colours, there are ${n \choose 3}$ possible choices for $r$, $s$, and $t$;
each of these choices could conceivably produce a different
$W^{(n)}_{[rst]}$.  Second, the eigenvalues of the $\hat{V^2}$ operator
are difficult to find; even those of the $W$-matrix are distinctly
non-trivial, since all that Eq.~(\ref{Weq3}) gives us is the matrix
entries.  Third, the matrix entries themselves, as given by Eq.
(\ref{Weq3}), are not easy to calculate analytically.

With these problems, it is to our advantage to make some
simplifications.  Instead of considering an arbitrary vertex, we will
examine the case of a monochromatic vertex for which all of the
matrices $W^{(n)}_{[rst]}$ are identical (since our ``graspings'' are
always on edges of the same colour).\footnote{In this case, we will
often suppress the indices $[rst]$, since they are always the same; in
other words, when referring to the monochromatic case, we will define
$W^{(n)} = W^{(n)}_{[rst]}$.} The formula for volume reduces to
\begin{equation}
\label{V2andW}
  \hat{V^2} = {n \choose 3} \left|\frac{iW^{(n)}}{16}\right| =
  \frac{n(n-1)(n-2)}{96} \left|iW^{(n)}\right|
\end{equation}

This solves our first problem.  The second problem, however, is
somewhat less tractable.  The matrix $W^{(n)}$ has a row for each
possible intertwiner core - usually a fairly large number.  To get
around this, we will examine bounds on the eigenvalues instead of the
eigenvalues themselves.  For any $n \times n$ matrix with entries
$a_{ij}$, any given eigenvalue $\lambda$ of this matrix satisfies
\begin{equation} \label{eiglim}
  | \lambda | \leq \max_i \sum_{j=1}^n |a_{ij}| \quad \mbox{and} \quad |
  \lambda | \leq \max_j \sum_{i=1}^n |a_{ij}|.
\end{equation}
In other words, the magnitude of every eigenvalue must be less than the
sum of the absolute values of some column and of some row.  Using
this, we can find an upper bound $M$ on the magnitudes of the
eigenvalues of $W^{(n)}$.  In the monochromatic case the eigenvalues
$\lambda_{W\alpha}$ of the $\hat{W}^{(n)}$ operator are related to the
eigenvalues $\lambda_{V^2\alpha}$ of the $\hat{V^2}$ operator by
\begin{equation}
  \lambda_{V^2\alpha} = \frac{n(n-1)(n-2)}{96} \left| \lambda_{W\alpha}
\right| \end{equation} as can be seen from (\ref{V2andW}).  Hence, we can
place a limit on the eigenvalues of the volume operator,
$\lambda_{V\alpha}  = \sqrt{\lambda_{V^2\alpha}}$, \begin{equation}
  \lambda_{V\alpha} \leq  \sqrt{ \frac{n(n-1)(n-2)}{96} M}. \label{Vlim}
\end{equation}

The determination of the upper bound $M$ is also rather complex.
Simpler expressions do exist for the case of a 4-valent vertex; in
this case, we can easily calculate the matrix elements explicitly.
For higher valence vertices, numerical techniques come in handy.

\subsection{Volume Eigenvalue Bounds for 4-valent Vertices}
\label{4vbounds}

In the case of a 4-valent vertex, De Pietri \cite{DPVol2} has shown
that the entries of the $W^{(4)}$ matrix for a 4-valent vertex with
edge colours $a$, $b$, $c$, and $d$ are given explicitly by the
formula
  \begin{eqnarray}
  W^{(4)}_{[012]} {}^{t+\epsilon}_{t-\epsilon} & = & \frac{-\epsilon
  (-1)^{(a+b+c+d)/2}}{32\sqrt{t(t+2)}} \nonumber \\
  & & \quad \left[ (a+b+t+3)(c+d+t+3)(1+a+b-t) \right. \nonumber \\
  & & \qquad (1+c+d-t)(1+a+t-b)(1+b+t-a) \nonumber \\
  & & \qquad \left. (1+c+t-d)(1+d+t-c) \right]^{\frac{1}{2}} \mbox{.}
  \label{4val2} \end{eqnarray}
where $t+\epsilon$ and $t-\epsilon$ correspond to the internal edge of the
vertex.  It can be shown that these elements are zero unless $\epsilon =
\pm 1$;  thus, each row or column will have at most two non-zero elements. 
To find $M$ in this case all we have to do is to find the maximum
value of this polynomial.

In the monochromatic case, $a=b=c=d$; we call this edge colour
$m$.  The polynomial in question becomes
\begin{equation}  \label{4lim1}
  \left|W^{(4)}_{[012]} {}^{t+\epsilon}_{t-\epsilon} \right| =
  \frac{(2m+t+3)(2m-t+1)(t+1)^2}{32\sqrt{t(t+2)}}.
\end{equation}
Thinking of $W$ as a continuous function $W(t)$, this expression is
maximized at 
\begin{equation} \label{4lim2} t_{\mathit{max}} = -1 +
\sqrt{\frac{(2m+2)^2 + 4}{6} + \frac{1}{6} \sqrt{(2m+2)^4 - 16(2m+2)^2
+ 16}}\mbox{,}
\end{equation}
and the maximum value is, of course, $W(t_{\mathit{max}})$.  In the
large-spin limit, where $m \gg 1$, this maximum value scales as
\begin{equation}
  \left| W_{\mathit{max}} \right| \approx \frac{m^3}{6\sqrt{3}}.
\end{equation}
The maximum absolute row or column sum (i.e. the quantity $M$) is at
most twice this number; therefore, (\ref{Vlim}) tells us that for a
4-valent monochromatic vertex with edge colour $m$, any eigenvalue
$\lambda_V$ of the volume operator on this vertex will satisfy (more
or less) the inequality
\begin{equation} \label{large4lim}
  \lambda_V \leq l_V^3 \frac{m^{3/2}}{\sqrt{12\sqrt{3}}}.
\end{equation}

This is a very encouraging result when we consider it in relation to
the area operator.  The eigenvalues of the area operator are given by
\cite{RSareavol,FLR,QGI} \begin{equation}
  \lambda_A = \ell_{P}^2 \sum_{i} \sqrt{m_i(m_i+2)}
\end{equation}
where the values $m_i$ are all the intersections of edges with the
surface in question.  For a 4-valent vertex with large edge colour
$m$, we have
\begin{equation}
  \lambda_A \propto m \mbox{.}
\end{equation}
Together with (\ref{large4lim}), this implies that the volume scales
no faster than $A^{3/2}$, which is what we would expect from classical
spatial geometry.

\begin{figure}
       \begin{center}
       \includegraphics[width=4in]{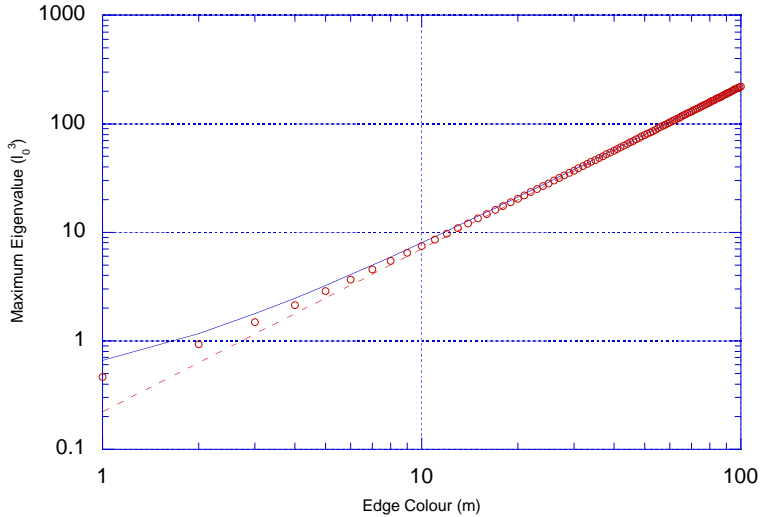}
       \caption{\label{4eigsandlim} The eigenvalue limit obtained in
       (\ref{4lim1}) and (\ref{4lim2}), compared with the maximum volume
       eigenvalue for $1 \leq m \leq 100$.  The lower dotted line is the
       result of a curve fit of the maximum eigenvalues to a curve of the
       form $\lambda = k m^{3/2}$, where $k$ is the only free parameter of
       the fit.  This fit yields a value of $k = 0.22155$.} \end{center}
\end{figure}

We must ask ourselves, however, how ``good'' this bound is: do the
eigenvalues actually scale proportionally to $m^{3/2}$, and if so, is the
proportionality coefficient we have found (i.e. $1/\sqrt{12\sqrt{3}}$)
close to the actual proportionality coefficient?  Figure \ref{4eigsandlim}
shows the largest eigenvalue of the volume operator (found by numerical
calculation) compared to the bound stated in (\ref{4lim1}) and
(\ref{4lim2}).  As we can see, the bound fits the volume eigenvalues very
tightly, diverging significantly from them for $n<10$ and becoming
practically indistinguishable from the maximum eigenvalue for $n>50$.

\subsection{Eigenvalue Bounds for $n$-Valent Vertices}

We now turn our attention to the more general case of the $n$-valent
monochromatic vertex.  In this case, we cannot use the simple formula of
the four-valent case, but must instead use the more general form. There
are some simplifications that can be made which aid us in computing these
matrix entries and their bounds.  Numerical methods prove to be fruitful.

We recall that we are examining the special case where $P_{r} = P_{s} =
P_{t} = P_{0} = m$.  We note that the expression for the 9-$j$ symbol in
Eq.  (\ref{Weq3}) is fairly symmetric, and can be simplified.  This symbol
corresponds to the spin network
\begin{equation}
  \left\{  \begin{array}{ccc}
  k_2 & P_t & k_3 \\
  i_2 & P_t & i_3 \\
  2 & 2 & 2
\end{array} \right\} = \eqngraph{Hex_Net2}.
\end{equation}
Using usual recoupling moves we can move the graspings of the 2-edges and
close the resulting ``triangles'' to obtain
\begin{eqnarray*}
  \left\{
  \begin{array}{ccc}
  k_2 & m & k_3 \\
  i_2 & m & i_3 \\
  2 & 2 & 2
  \end{array} \right\} & = & \frac{{\lambda}_{k_3}^{i_3 2}}{m} \left(
  \frac{-i_2 \, \mbox{Tet} \left[ \begin{array}{ccc} k_2 & i_2 & i_2 \\ 2
  & 2 & 2 \end{array} \right]}{\theta(k_2,i_2,2)} \right. \\
    & & \quad + \left. \frac{i_3 \, \mbox{Tet} \left[
  \begin{array}{ccc}
  k_3 & i_3 & i_3 \\
  2 & 2 & 2
  \end{array} \right]}{\theta(k_3,i_3,2)} \right) \mbox{Tet} \left[
  \begin{array}{ccc}
  k_2 & k_3 & 2 \\
  i_3 & i_2 & m
  \end{array} \right].
\end{eqnarray*}
Plugging this back in, (\ref{Weq3}) becomes
\begin{eqnarray*}
  \lefteqn{W^{(n)}_{[rst]} {}^{k_2 \cdots k_{n-2}}_{i_2 \cdots
  i_{n-2}} = {}} \\
  & & \quad - m^{3}
  {\mathcal Q} \lambda_{k_2}^{i_2 2} \delta_{i_4}^{k_4}
  \cdots \delta_{i_{n-2}}^{k_{n-2}}\nonumber \\
  & & \quad \, \times \frac{\mbox{Tet} \left[ \begin{array}{ccc}
  m & m & m \\
  k_2 & i_2 & 2
  \end{array} \right] \mbox{Tet} \left[ \begin{array}{ccc}
  m & m & k_4 \\
  k_3 & i_3 & 2
  \end{array} \right] \mbox{Tet} \left[
  \begin{array}{ccc}
  k_2 & k_3 & 2 \\
  i_3 & i_2 & m
  \end{array} \right]\Delta_{k_2} \Delta_{k_3}}
  {\theta(k_2, i_2, 2) \theta(k_3, i_3, 2) \theta(m, m, k_2)
  \theta(k_2, k_3, m) \theta(k_3, k_4, m)}
\end{eqnarray*}
where
\begin{equation}
  {\mathcal Q} = \frac{{\lambda}_{k_3}^{i_3 2}}{m} \left(
  \frac{-i_2 \, \mbox{Tet} \left[
  \begin{array}{ccc}
  k_2 & i_2 & i_2 \\
  2 & 2 & 2
  \end{array} \right]}{\theta(k_2,i_2,2)} +  \frac{i_3 \, \mbox{Tet}
  \left[ \begin{array}{ccc} k_3 & i_3 & i_3 \\ 2 & 2 & 2 \end{array}
  \right]}{\theta(k_3,i_3,2)} \right). \label{Qdef}
\end{equation}

We can write the Tets in this equation in terms of Wigner
6-$j$ symbols. At least one of the arguments of the resulting Wigner
6-$j$ symbols will be 1, and analytic formulas for this case have been
compiled (see, for example, Varshalovich \cite{Varsh}).  Equation
(\ref{Weq3}) then becomes \begin{equation}
  W^{(n)}_{[rst]} {}^{k_2 \cdots k_{n-2}}_{i_2 \cdots
  i_{n-2}} = - m^{3}
  {\mathcal Q R S} \lambda_{k_2}^{i_2 2} \delta_{i_4}^{k_4}
  \cdots \delta_{i_{n-2}}^{k_{n-2}}
\end{equation}
where $\mathcal{Q}$ is given in (\ref{Qdef}), and $\mathcal{R}$ and
$\mathcal{S}$ are given by
\begin{equation}
       {\mathcal R} = \left\{ \begin{array}{ccc}  m/2 & k_{2}/2 & k_{3}/2
       \\ 1 & i_{3}/2 & i_{2}/2 \end{array} \right\}_{W} \left\{
  \begin{array}{ccc}
       m/2 & m/2 & k_{2}/2 \\ 1 & i_{2}/2 & m/2 \end{array} \right\}_{W}
       \left\{ \begin{array}{ccc} k_{4}/2 & m/2 & k_{3}/2 \\ 1 & i_{3}/2 &
       m/2 
\end{array}
       \right\}_{W}
\end{equation}
\begin{equation}
       {\mathcal S} = \theta(m,m,2) \sqrt{\frac{\theta(i_{2}, i_{3}, m)
       \theta(m, m, i_{2}) \theta(i_{3}, i_{4}, m)}{\theta(k_{2}, k_{3},
       m) \theta(m, m, k_{2}) \theta(k_{3}, k_{4}, m)}}.
\end{equation}
Note that we are using a $W$ subscript to denote where we have used
Wigner 6-$j$ symbols instead of the usual spin-network 6-$j$ symbols; see
Ref.\ \cite{mikethesis} for the exact correspondence between these sets of
symbols.

Finally, we note that there are certain ``selection rules'' for
finding the non-zero elements of $W$ in this basis.  The presence of a
series of Kronecker deltas in (\ref{Weq3}) gives us the obvious selection
rules that
\begin{equation} \label{latersame}
  k_4 = i_4, \quad k_5 = i_5, \quad \cdots \quad k_{n-2} = i_{n-2}.
\end{equation}
It is also evident from the spin diagram in (\ref{9jdef}) that for the
9-$j$ symbol in (\ref{Weq3}) to be non-zero, we must have \begin{equation}
\label{2and3}
k_2 - i_2 = 0, \pm 2 \quad \mbox{and} \quad k_3 - i_3 = 0, \pm 2
\end{equation} in order to satisfy the triangle inequalities. 
Finally, it can be shown that \begin{equation} \frac{a}{\theta(a,b,2)}
\mbox{Tet} \left[ \begin{array}{ccc} a & a & b \\
  2 & 2 & 2 \end{array} \right] = \left\{ \begin{array}{l @{\quad} c}
  (-1)^{a+1} \cdot \frac{a+2}{2} & b = a-2 \\ -1 & b=a \\ (-1)^a \cdot
  \frac{a}{2} & b=a+2 \\ 0 & \mbox{otherwise} \end{array} \right.
\end{equation}
Hence, if $k_2 = i_2$ and $k_3 = i_3$, the two terms in parentheses in
(\ref{Qdef}) will be be equal, and therefore cancel out, making the
quantity $\mathcal{Q}$ zero.  We can then rewrite (\ref{2and3}) as
\begin{equation} \label{2and3rev}
  \begin{array}{c}
  k_2 - i_2 = 0, \pm 2, \quad  k_3 - i_3 = 0, \pm 2, \quad \mbox{with} \\
  k_2 - i_2 \mbox{ and } k_3 - i_3 \mbox{ are not both zero.} \end{array}
\end{equation}
We conclude that for a given column of the $W$ matrix (i.e. fixed
$k_2, k_3, \ldots,$ $k_{n-2}$), there will be at most eight non-zero
elements:
\begin{equation}
      \begin{array}{c @{\qquad} c @{\qquad} c}
      i_{2} = k_{2} - 2 & i_{2} = k_{2} - 2 & i_{2} = k_{2} - 2 \\
      i_{3} = k_{3} - 2 & i_{3} = k_{3} & i_{3} = k_{3} + 2 \\[10pt]
      i_{2} = k_{2} & & i_{2} = k_{2} \\
      i_{3} = k_{3} - 2 & & i_{3} = k_{3} + 2 \\[10pt]
      i_{2} = k_{2} + 2 & i_{2} = k_{2} + 2 & i_{2} = k_{2} + 2 \\
      i_{3} = k_{3} - 2 & i_{3} = k_{3} & i_{3} = k_{3} + 2
      \end{array}
\end{equation}

The quantities $\mathcal{Q}$, $\mathcal{R}$, and $\mathcal{S}$ are
fairly simple to calculate for a given set of arguments.  Finding the
maximum product of all three, over all possible values of $k_2, k_3, k_4,
i_2, i_3$, and $i_4$, is a much more daunting task.\footnote{ Note that,
in a small mercy granted to us by equation (\ref{Weq3}), the entries of
this form of the $W$-matrix depend only on these six intertwiner strands
(and the edge colour $m$) and not on the entire set.} This task is further
complicated by the fact that both the quantities $\mathcal{Q}$ and
$\mathcal{R}$ have slightly different analytic forms for each of the eight
entries in a given column --- one form for $k_2 - i_2 = 2$ and $k_3 - i_3
= 2$, one form for $k_2 - i_2 = 2$ and $k_3 - i_3 = 0$, and so forth (see
Varshalovich \cite{Varsh}.)

Despite this difficulty, we can get an idea of the scaling properties of
the absolute row and column sums by numerical analysis.  By a happy
coincidence, we do not even need to look at all of the rows and columns of
the matrix --- only a small subset of them.  This is because (as mentioned
above) the entries of the $W$-matrix are only dependent on the six values
$\{k_2, k_3, k_4, i_2, i_3, i_4\}$; all other pairs of corresponding
indices must be equal, or the matrix entry will be zero.  These facts
imply that the $W$-matrix must be of block-diagonal form, and, more
importantly, that each one of the blocks is identical.  We see that we
need only compute the absolute row and column sums for the (much smaller)
block matrix, instead of the full $W$-matrix.  This is still a non-trivial
calculation, as the size of this matrix scales fairly quickly (for $m =
10$ this matrix has already grown to $891 \times 891$) but it does
simplify the problem greatly.

\begin{figure}
       \begin{center}
       \includegraphics[width=4in]{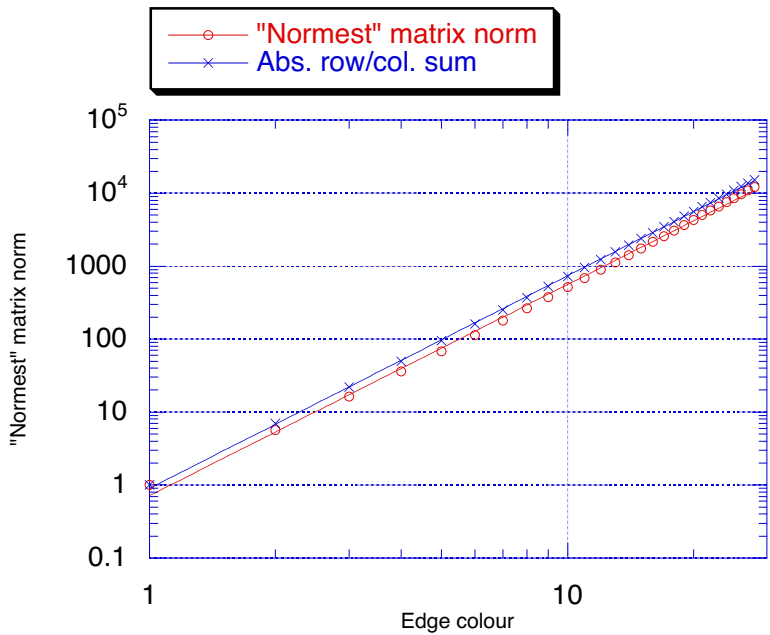}
       \caption{\label{Mvals} The eigenvalue limits found by the absolute
       row and column sums and by MATLAB's internal \ttfamily normest
       \rmfamily function.  The curve fits are for a power law relation. 
       For the absolute row/column sum method, the best fit is $M =
       0.90432 \cdot x^{2.9126}$;  for the \ttfamily normest \rmfamily
       method, it is $M = 0.71884 \cdot x^{2.8921}$.} \end{center}
\end{figure}

Using this trick, combined with the selection rules in
(\ref{latersame}) and (\ref{2and3rev}), the bounding quantity $M$ (as
defined above in section \ref{4vbounds}) may be calculated.  Two methods
were used.  The first was the absolute row and column sum discussed
previously.  The second was MATLAB's internal \ttfamily normest \rmfamily
function, which estimates the largest eigenvalue of a sparse
matrix.\footnote{ For the mathematically inclined reader, we note that the
absolute row and column sums are known as the 1-norm and the $\infty$-norm
of a given matrix, respectively; that which \ttfamily normest \rmfamily
finds is the 2-norm.} These results are shown in Figure \ref{Mvals}.  Each
of these data sets is fitted to a power law relation.  We see that for
both of these data sets, we have an approximate relation $M \propto m^3$.

There is good reason to believe that this power dependence should tend
towards $m^3$ (exactly) as $m \to \infty$: if it does, then we have the
relation
  \begin{eqnarray}
  \left| \lambda_V \right| & \approx & \sqrt{M \frac{n(n-1)(n-2)}{96}}
  \nonumber \\ & \approx & C \sqrt{m^3 n^3} \nonumber \\ & = & C
  \mbox{(total spin)}^{3/2} \end{eqnarray}
where $C$ is a proportionality constant.  As noted in the previous
section, this is the dependence expected from classical spatial
geometry.  While our data are not conclusive evidence of this
relation, they are highly suggestive that this relation holds for a
monochromatic vertex of any valence.

\section{Conclusions}
\label{conc}

We have seen some interesting consequences stemming from the
discreteness of the angle operator.  These results will be referred to as
the \emph{small angle} property, the \emph{angular resolution} property,
and the \emph{angular distribution} property.  We now wish to assign some
scaling behavior to these predicted phenomena; in other words, on what
length scales do we expect these phenomena to be observable?

The small angle property is the best-quantified of these three
properties. We have the equation
  \begin{equation}
  \epsilon \geq \frac{4}{\sqrt{n + 8}} \geq \frac{4}{\sqrt{s_{T} + 8}}
  \mbox{,} \end{equation}
where $s_{T}$ is the total surface flux.  From our numerical studies
of the volume operator, we also have the result that
\begin{equation} \label{len2spin}
  V \propto L^{3} \propto s_{T}^{3/2}
\end{equation}
If we combine these results then we obtain the result that for
large $s_{T}$, the minimum observable angle is related very simply to the
length scale observed \begin{equation}
  \epsilon \propto L^{-1} \propto s_{T}^{-1/2}
\end{equation}
where $L$ is the characteristic length (in Planck units) defined by
the volume or area.  Thus, an accuracy of $10^{-10}$ radians
corresponds to spin fluxes of $10^{20}$ and spin network vertices of
$L \sim 10^{-25}$m.  Moreover for large ``background geometries'' the
primary limit is not from $s_{T}$ but from $(s_{1}+s_{2})$ \[
  \epsilon \propto (s_{1}+s_{2})^{-1/2}.
\]

It is also possible to directly relate the minimum angle to the volume
contribution of the spin network vertex.  Using the same relations as
above the dependence of minimum angle and volume on the total spin flux
implies the ``minimum-angle volume bound'' given by
  \begin{equation}
      \epsilon \frac{V^{1/3}}{\ell_P} \gtrsim 1.
  \end{equation}
One way to read this is that the smaller the minimum angle desired, the
larger the volume contribution of the vertex, the larger the ``quanta of
volume.''  Thus, this particular form of the angle operator suggests a
very different description of our flat space.  The effectively continuous
spin network state would be a highly flocculent network of high valent
vertices.

We note that this ``bound'' is based only on the considerations of the
dependence of the spectra on the spin flux.  It is not derived by the
non-commuting properties of the operators.  In addition to this bound, we
expect that there would be a volume-angle uncertainty relation as there is
in the case of the area operators \cite{QGIII}.

In Section \ref{angres} we found results for the mean angular
resolution associated with a vertex.  The most important of these were
the average spacing between expectation values for momochromatic
vertices, given by
  \begin{equation}
  \delta \geq \frac{4 \pi}{n^2 + 3n + 3} \mbox{.}
  \end{equation}
Combining these results with the result in Eq.\ (\ref{len2spin})
allows us to associate a scaling behavior with the mean angular
separation and resolution.  For large $n$, the angular separation
scales as
  \begin{equation}
      \delta \propto n^{-2} \propto L^{-4} \mbox{.}
  \end{equation}
This scaling property of the angular resolution is somewhat striking,
since it scales proportionally to such a high value of the total spin
(i.e. inversely proportional to the fourth power of the length.)

We found that we could reproduce the classical angle distribution of
${\cal P}(\theta) \, \mathrm{d} \theta = \sin \theta \, \mathrm{d}
\theta$ under the condition that
\[
  1 \ll \{ s_{1}, s_{2} \} \ll s_{r}.
\]
In the absence of good analytical control on the precise behavior, it is
not possible to determine the length scale to this condition.  The
importance of our angle distribution results is more clear in light of the
small-angle results.  The minimum angle observable depends on $s_{T}$ if
$s_{1}, s_{2} \approx s_{r}$, but on $s_{1}+s_{2}$ if $s_{1}, s_{2} \ll
s_{r}$.  To obtain the classical distribution the latter case holds and so
the sum of the fluxes $s_{1}$ and $s_{2}$ must also be large.  Thus, to
approximate classical angles the total spin and the spin through surface
patches $S_{1}$ and $S_{2}$ must be quite large.

Finally, in reflecting on these results it is important to keep in
mind the nature of angle operator under consideration.  Classically an
angle is defined at the intersection of two lines - a point.  In loop
quantum gravity it is possible to define the angle at a single spin
network vertex.  As the above analysis shows, these ``atoms of
geometry'' are small objects.  It seems likely that to model familiar
interactions which have ``angle dependence'' it will be necessary to
average over many such vertices.  If this in fact proves to be the
case, then the above estimates must be modified.

\begin{ack}
It is a pleasure to acknowledge the generous hospitality of Deep
Springs College and summer research funds from Swarthmore College.
\end{ack}

\end{document}